\documentclass[conference, 11pt,letter]{IEEEtran}
\IEEEoverridecommandlockouts
\usepackage{romannum}
\usepackage{amsmath}
\usepackage{graphicx}
\usepackage{color}
\usepackage{multirow}
\usepackage{amssymb,bm}
\usepackage{algpseudocode}
\usepackage{algorithm}
\usepackage{array}
\usepackage{enumerate}
\usepackage{float}
\usepackage{subfigure}
\usepackage{amsthm}
\usepackage{thmtools}
\usepackage[utf8]{inputenc}
\usepackage[T1]{fontenc}
\usepackage{url}
\usepackage{ifthen}
\usepackage{cite}
\usepackage{mathtools}
\usepackage{bbm}
\usepackage{braket}

\declaretheoremstyle[%
spaceabove=6pt,%
spacebelow=6pt,%
headfont=\normalfont\itshape,%
notefont=\normalfont\itshape,
qed=\qedsymbol,%
headpunct={:},
bodyfont=\normalfont,%
]{mystyle}

\declaretheoremstyle[%
spaceabove=6pt,%
spacebelow=6pt,%
headfont=\bfseries,%
bodyfont=\normalfont,%
postheadspace=1em,%
]{mystyle_1}

\makeatletter
\newcounter{phase}[algorithm]
\newlength{\phaserulewidth}

\makeatother

\begin{document}
\bstctlcite{IEEEexample:BSTcontrol}

\pagenumbering{arabic}
\graphicspath{{./resources/}}

\title{Variational Quantum Compressed Sensing for \\ Joint User and Channel State Acquisition \\
in Grant-Free Device Access Systems}

\author{Bryan Liu\IEEEauthorrefmark{1}\IEEEauthorrefmark{2}\thanks{B. Liu conducted this research during his internship at MERL.}, Toshiaki Koike-Akino\IEEEauthorrefmark{1}, Ye Wang\IEEEauthorrefmark{1}, 
	Kieran Parsons\IEEEauthorrefmark{1} \\

\IEEEauthorblockA{\small\IEEEauthorrefmark{1}Mitsubishi Electric Research	Laboratories (MERL), 201 Broadway, Cambridge, MA 02139, USA.}

\IEEEauthorblockA{\small\IEEEauthorrefmark{2}Electrical Engineering and Telecommunication, University of New South Wales, Sydney, Australia.}
	Email: bryan.liu@unsw.edu.au, \{koike, yewang, parsons\}@merl.com
}
\maketitle

\begin{abstract}
    This paper introduces a new quantum computing framework integrated with a two-step compressed sensing technique, applied to a joint channel estimation and user identification problem. 
    We propose a variational quantum circuit (VQC) design as a new denoising solution. 
    For a practical grant-free communications system having correlated device activities, variational quantum parameters for Pauli rotation gates in the proposed VQC system are optimized to facilitate to the non-linear estimation. 
    Numerical results show that the VQC method can outperform modern compressed sensing techniques using an element-wise denoiser.
\end{abstract}


\section{Introduction}

With the increasing number of devices that are connected to wireless communications systems, grant-free access schemes have been proposed to accommodate massive connectivity.
A grant-free access scheme allows an unspecified number of devices to transmit data instantaneously whenever they have a transmission request.
Considering the fact that the device activity pattern is often sparse, compressed sensing (CS) algorithms can achieve an outstanding performance on channel estimation and device identification in a massive connectivity channel~\cite{cs_mMTC}.
State-of-the-art CS techniques such as approximate message passing (AMP)~\cite{amp}, vector approximate message passing~\cite{vamp}, and orthogonal approximate message passing (OAMP)~\cite{oamp} are used to jointly recover the channel coefficients and device activity.
These CS algorithms perform a two-step iterative process, which alternates linear estimation (LE) and non-linear estimation (NLE) steps.
The LE step returns an estimated sequence, where the elements can be modeled by independent and identically distributed (\textit{i.i.d.}) variables that are corrupted by additive white Gaussian noise (AWGN).
The NLE step can be performed element-wise and a state evolution can be derived to theoretically predict the convergence performance of the recovery process~\cite{gamp}.
However, an element-wise NLE function is constructed based on the assumption that each device's activity is independent.
In a massive connectivity network, the activities are generally correlated, which contradicts the assumption of classical CS algorithms.
In this paper, we investigate a variational quantum circuit (VQC) that applies quantum gates to address device activity correlation to improve the compressed sensing performance. 

Quantum computation has shown its promising potential in the development of cryptography~\cite{quantum_cryptography}, information theory~\cite{quantum_it}, physics~\cite{feynman_quantum}, and mathematics~\cite{Shor_alg}. 
For instance, Shor's algorithm can achieve an exponential speedup over the classical algorithms for prime factorization of integers for fault-tolerant quantum processors. 
More recently, variational quantum algorithms~\cite{qaoa, qaoa_ml, ml_ai_quantum} have been proposed to be robust against quantum decoherence in noisy intermediate-scale quantum (NISQ) processors~\cite{bharti2021noisy}.
It was reported in~\cite{arute2019quantum, zhong2020quantum} that \emph{quantum supremacy} has been achieved with a real quantum processor for some specific problems.
Accordingly, quantum computing is envisioned as a key driver for the sixth generation (6G) applications~\cite{qml_survey}

In this paper, we consider the channel estimation and user identification in a sparse activity system by using a variational quantum algorithm. 
To explore the correlation structure, we propose a new VQC-based denoiser in the NLE step of the CS algorithm. 
Numerical results show that our VQC ansatz is able to outperform the element-wise soft-thresholding and minimum mean-square error (MMSE) denoisers. 
The key contributions of this paper are summarized as follows:
\begin{itemize}
    \item We investigate practical correlated device activities in grant-free access systems.
    \item We propose a quantum-based compressed sensing to identity device activity and channel state jointly.
    \item We introduce a computationally efficient ansatz for VQC to denoise correlated signals.
    \item We further adopt post-processing deep neural network to improve the device detection accuracy.
    \item We validate the performance of VQC-based denoising and post processing approach.
\end{itemize}
To the best of authors' knowledge, this paper is the very first paper employing quantum computing for CS.

\section{Grant-Free IoT-Device Access Systems}

\subsection{System Model} 

\begin{figure}[t]
    \centering
    \includegraphics[width=0.95\linewidth]{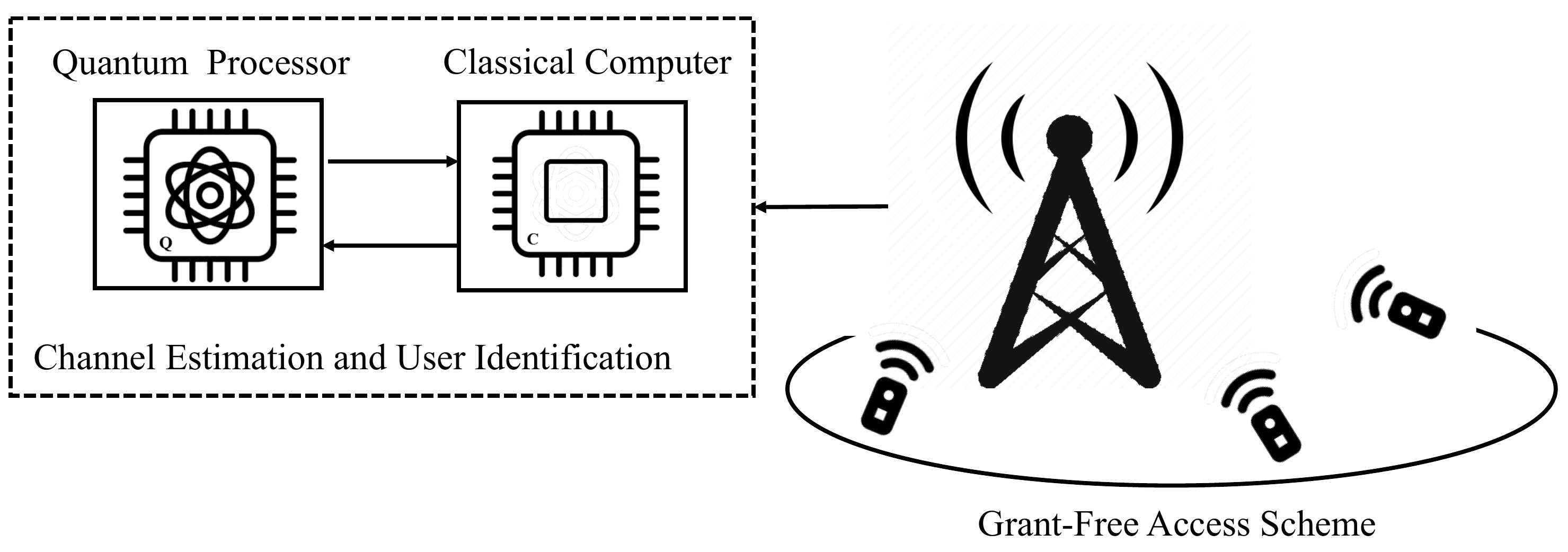}
    \caption{Grant-free access systems empowered by a classical-quantum hybrid processing for data/device detection.}
    \label{Channel_model_plot}
\end{figure}

Fig.~\ref{Channel_model_plot} shows grant-free wireless access systems having unspecified number of devices to communicate with a base station which is empowered by a quantum processor (such as the IBM quantum cloud computing).
Following most typical compressed sensing problems, we consider a linear recovery formulation:
\begin{align}\label{system_model}
    \mathbf{y} = \mathbf{A}\mathbf{x} + \mathbf{z},
\end{align}
where $\mathbf{y} \in \mathbb{C}^{M}$ represents $M$ received symbols, $\mathbf{A} \in \mathbb{C}^{M\times N}$ represents a pilot matrix for $N$ devices with $M < N$, and $\mathbf{z} \stackrel{\text{iid}}{\sim} \mathcal{CN}( \mathbf{0}, \sigma^2 \mathbf{I})$ represents the AWGN of noise variance $\sigma^2$ with $\mathbf{0}$ and $\mathbf{I}$ begin all-zeros vector and identity matrix of size $M$, respectively.
The term $\mathbf{x} = \mathbf{a} \cdot \mathbf{h}$ in \eqref{system_model}
indicates an unknown device activity $\mathbf{a} \in \mathbb{F}_2^N$, distorted by a fading channel coefficient $\mathbf{h} \in \mathbb{C}^N$.
While this channel model corresponds to a simplified version of practical narrow-band Internet of things (NB-IoT) systems~\cite{ce_NBIOT}, extending to wide-band scenarios is relatively straightforward.

Although most CS works assume \textit{i.i.d.} device activity, typical grant-free network may face a correlated user activity due to shared medium environment.
To model such a correlated access network, the elements $a_i$ and $a_j$ for $i, j \in \{1, 2, \ldots, N\}$ in the device activity pattern $\mathbf{a}$ are assumed to be autoregressively correlated by a correlation coefficient $\gamma^{|i-j|}$. 
We assume that the correlation coefficient is unknown to our VQC-based compressed sensing (denotes as VQC-CS) and we expect the VQC explores the correlation structure during the learning phase, which is controlled by a classical computer as shown in Fig.~\ref{Channel_model_plot}.

\subsection{Compressed Sensing: OAMP Algorithm}

As a state-of-the-art CS technique, we briefly describe the OAMP algorithm~\cite{oamp}.
The OAMP algorithm can usually achieve a better convergence performance than conventional CS techniques, such as the fast iterative soft-thresholding algorithm (FISTA) and AMP algorithm~\cite{amp}.
The OAMP algorithm inherits the two-step iterative processes, consisting of a decorrelated LE step and a divergence-free NLE step. 
The iterative process can be summarized as: 
\begin{align}
    &\text{(LE):} \ & \mathbf{l}^t &= \hat{\mathbf{x}}^t + \mathbf{D}(\mathbf{y} - \mathbf{A}\hat{\mathbf{x}}^t), 
    \label{eq:le}
    \\
    &\text{(NLE):} \ & \hat{\mathbf{x}}^{t+1} &= p_t\bigg{
    (}\eta_t(\mathbf{l}^t) - \Big{
    (}\frac{1}{N}\sum_{i=1}^N\eta_t'(l_i^t) \Big{
    )} \mathbf{l}^t 
    \bigg{)},
    \label{eq:nle}
\end{align}
where $\eta_t(\cdot)$ is a denoiser function, $\eta_t'(\cdot)$ is its first-order derivative, and $p_t$ is a scaling factor that can be optimized at the $t$th iteration~\cite{oamp}. 
Based on the pilot matrix $\mathbf{A}$, the matrix $\mathbf{D}$, which decorrelates the elements in the linear estimate, can be found by
$\mathbf{D} = \frac{N}{\text{tr}\{\hat{\mathbf{D}}\mathbf{A}\}}\hat{\mathbf{D}}$, where $\hat{\mathbf{D}}$ could be in the form of matched filtering, pseudo-inverse, or linear MMSE:
$\hat{\mathbf{D}}_{\text{MF}} = \mathbf{A}^\text{H}$;
$\hat{\mathbf{D}}_{\text{PINV}} = \mathbf{A}^\text{H}(\mathbf{A}\mathbf{A}^\text{H})^{-1}$; 
$\hat{\mathbf{D}}_{\text{LMMSE}} = \tau^2\mathbf{A}^\text{H}(\tau^2\mathbf{A}\mathbf{A}^\text{H} + \sigma^2\mathbf{I})^{-1}$ where
$\tau^2$ is an empirical MSE of the NLE. 
The OAMP alternating process maintains the orthogonality between the estimation errors of LE and NLE, achieving an outstanding convergence performance for solving the linear recovery problem in general.

\subsection{Quantum Computing: Variational Quantum Circuit}

Quantum computing leverages quantum physics phenomena, such as superposition and entanglement, to yield over classical computers~\cite{qa_book}.
The superposition $\ket{\psi}$ of a qubit can be represented by an orthonormal basis vector $\ket{0}=[1, 0]^\text{T}$ and $\ket{1} = [0, 1]^\text{T}$ as $\ket{\psi} = q_0 \ket{0} + q_1\ket{1}$ with $|q_0|^2 + |q_1|^2 = 1$.
The Hermitian transpose of $\ket{\psi}$ is denoted as $\bra{\psi}$. 
In convention, $\ket{\psi}$ can be represented by the Bloch sphere with
$q_0 = \cos(\theta/2)$ and $q_1 = \mathrm{e}^{i\varphi}\text{sin}(\theta/2)$,
where $\theta \in [0, \pi]$ and $\varphi \in [0, 2\pi]$ represent the latitude and longitude of the sphere, respectively.
Rotation gates $R_X$, $R_Y$ and $R_Z$ in a quantum circuit over the $X$, $Y$, and $Z$-axes are defined as $R_X = \bigl[ \begin{smallmatrix}\text{cos}(\theta/2) & -i\text{sin}(\theta/2)\\ -i\text{sin}(\theta/2) &  \text{cos}(\theta/2) \end{smallmatrix}\bigr]$, $R_Y = \bigl[ \begin{smallmatrix}\text{cos}(\theta/2) & -\text{sin}(\theta/2)\\ \text{sin}(\theta/2) &  \text{cos}(\theta/2) \end{smallmatrix}\bigr]$ and $R_Z = \bigl[ \begin{smallmatrix}e^{-i\theta/2} & 0 \\ 0 &  e^{i\theta/2} \end{smallmatrix}\bigr]$, respectively. 

A VQC refers to the quantum circuit that is specified by variational parameters. 
For instance, the rotation angles of the rotation gates can be calibrated according to a preset cost function. 
Similar to widely used deep neural networks for channel estimation~\cite{ce_dnn, ce_NBIOT}, the parameterized rotation angles for VQC are determined through a learning phase. 
With the so-called parameter-shift rule~\cite{schuld2019evaluating}, VQC can be analytically differentiable, that enables stochastic gradient optimization of variational parameters likewise DNNs.
In this paper, we investigate a full system that embeds a VQC into the compressed sensing technique, where the VQC behaves as a scaling factor processor to adjust the estimate before feeding back to the linear estimator.

\section{Variational Quantum Compressed Sensing}

\subsection{VQC-based Compressed Sensing}

\begin{figure}[t]
    \centering
    \includegraphics[width=\linewidth]{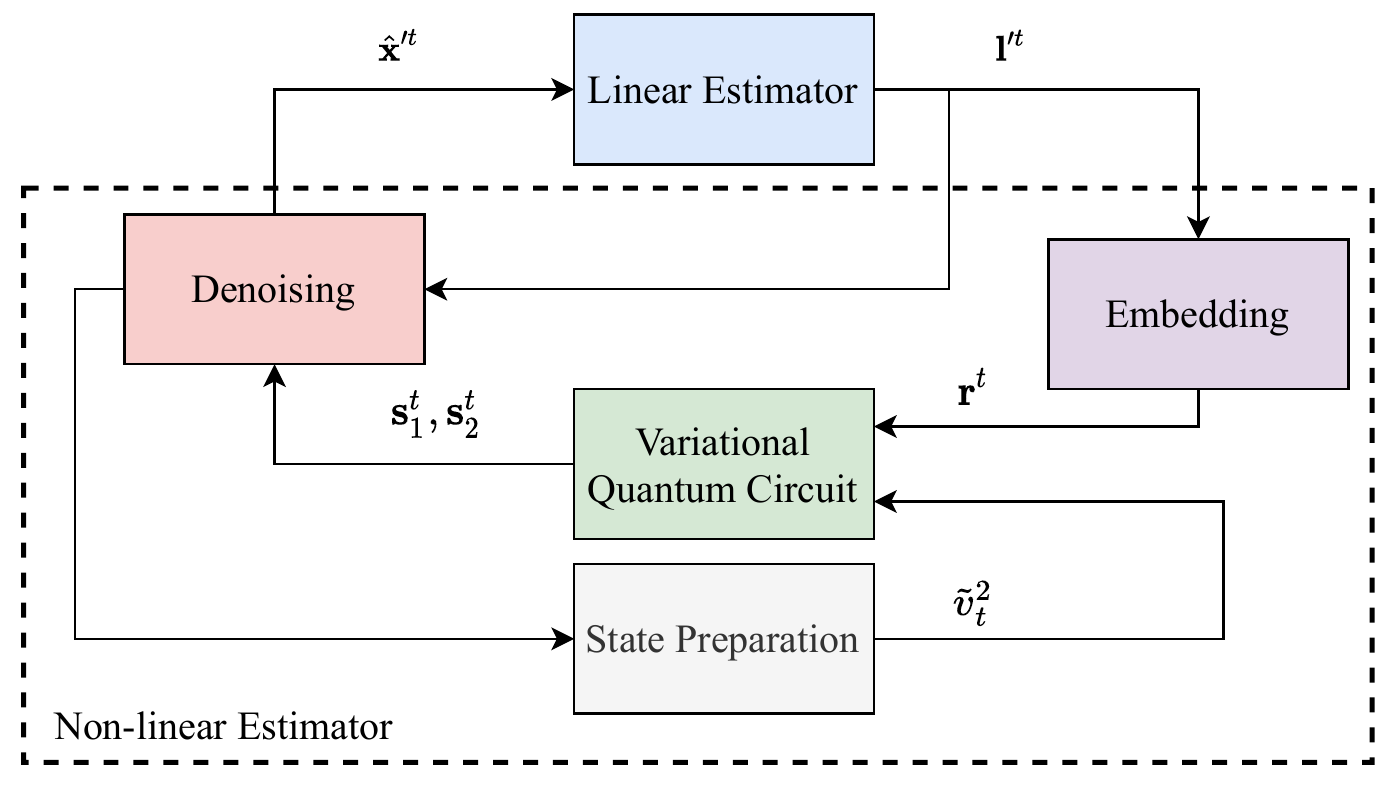}
    \caption{Block diagram of the proposed VQC-CS.}
    \label{VQC_CS_block}
\end{figure}

The main motivation of the proposed VQC-CS is to integrate a quantum circuit into a compressed sensing algorithm so that the correlation between the device activities can be exploited by the VQC-based denoising in the NLE step. 
Fig.~\ref{VQC_CS_block} shows a schematic of the proposed VQC-based compressed sensing, whose NLE step comprises $4$ sub-processes: i) embedding; ii) scaling; iii) error estimation; and iv) VQC.

\subsubsection{\textbf{Embedding}}
We embed the LE estimate $\mathbf{l}^t$ in \eqref{eq:le} into rotation angles of the VQC by limiting the value range in $(0, \pi)$ as follows:
\begin{align}
    r^t_i=\pi\cdot \tanh\big(
    |l_i^t|^2
    \big), 
    \label{r_embedding}
\end{align}
for $i \in \{1, 2, \ldots, N\}$. 

\subsubsection{\textbf{Denoising}}
The LE estimate is refined by a denoising function in NLE step, which employs a scaling operation.
Specifically, the $i$th device's NLE estimate in $\mathbf{x}^t$ for $i \in \{1, 2, \ldots, N\}$ is computed as: 
\begin{align}
\hat{x}^t_i 
&= \frac{s_{1, i}^t }{1+s_{2,i}^t} l^t_i,
\label{scaling} 
\end{align}
where ${s}^t_{1,i}$ and ${s}^t_{2,i}$ are scaling factors determined by 2 VQCs. 
Our introduction of the two scaling factors, $\mathbf{s}^t_1 \in \mathbb{R}^N$ and $\mathbf{s}^t_2 \in \mathbb{R}^N$, are motivated by the analogous form of MMSE denoiser~\cite{oamp, cs_mMTC}, \textcolor{black}{to stabilize the performance by having more measurements from the VQCs than using a single scaling factor.
Here we assume that the scaling factors $\mathbf{s}^t_1$ and $\mathbf{s}^t_2$ are the average measurements from 2 separate VQCs. 
The final denoising scale $\frac{s_{1, i}^t }{1 + s_{2,i}^t}$ is determined after obtaining the averaged measurements $s_{1, i}^t $ and $s_{2,i}^t$ for $i \in \{1, 2, \ldots, N\}$.} 

\subsubsection{\textbf{State Preparation}}
In order to feed the information of estimation error into the VQC, at each iteration we compute the recovery error of the received symbols as:
\begin{align}
    \tilde{v}_t^2 &=  
    \pi\cdot \tanh\Big(
    \frac{1}{N}
    \| 
    \mathbf{y} - \mathbf{A} \hat{\mathbf{x}}^t
    \|^2
    \Big).
    \label{recovery_ee_embed}
\end{align}
Similar to the embedding model for the LE estimate in \eqref{r_embedding}, the empirical recovery error is converted in a range of $0 \leq \tilde{v}_t^2 \leq \pi$, where $\tilde{v}_t^2$ is used to initialize the quantum state of the VQC as follows.

\begin{figure*}[t]
    \centering
    \includegraphics[width=\linewidth]{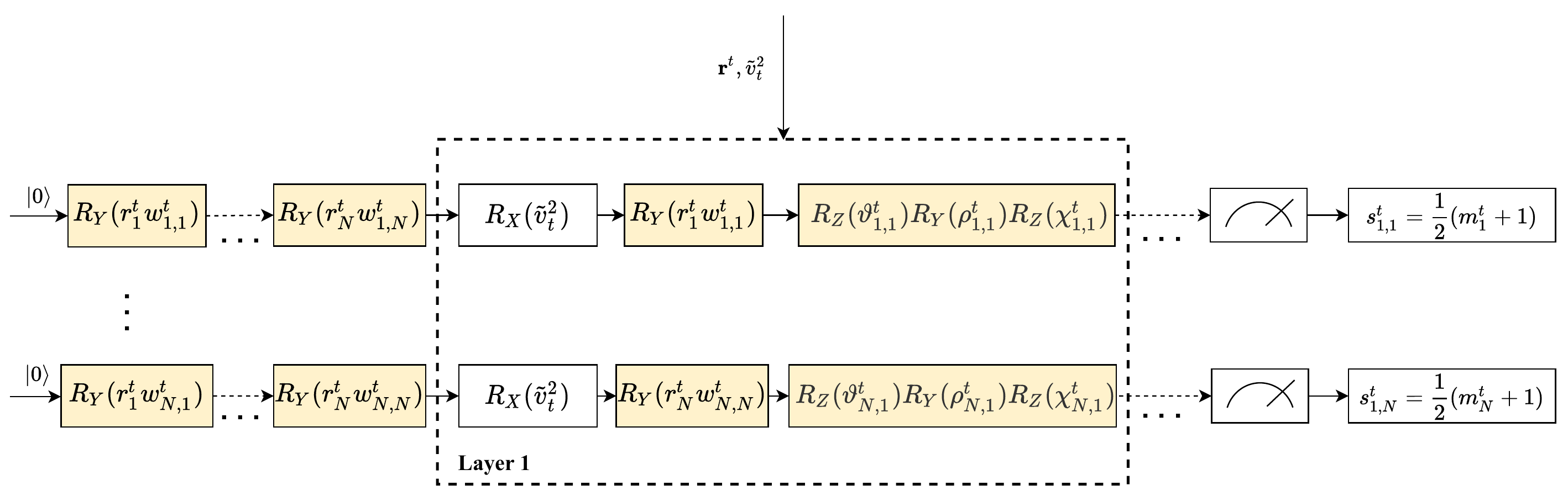}
    \caption{VQC ansatz for compressed sensing, with data re-uploading layers to generate the scaling factors for denoising.}
    \label{VQC}
\end{figure*}

\subsubsection{\textbf{Variational Quantum Circuit}}
The embedding information $\mathbf{r}^t$ and $\tilde{v}_t^2$ will serve as the input for the VQC, which in turn provides the scaling factors $\mathbf{s}_1^t$ and $\mathbf{s}_2^t$ for denoising. 
Fig.~\ref{VQC} illustrates our quantum gate ansatz, which is used in two individual VQCs to return the two scaling factors. 
The VQC uses trainable parameters $\mathbf{w}^t \in \mathbb{R}^{N \times N}$ to scale the input $\mathbf{r}^t$ before feeding into the $Y$-rotation gate. 
The shaded blocks refer to the rotation gates that contain such a tunable parameter to adjust the rotation angle. 
$\bm{\vartheta}_i^t \in \mathbb{R}^L$, $\bm{\rho}^t_i \in \mathbb{R}^L$ and $\bm{\chi}_i^t \in \mathbb{R}^L$ are the rotation angles of three concatenated $Y$-rotation, $Z$-rotation and $Y$-rotation gates in the $i$th qubit. 
$L$ indicates the number of layers in the VQC. 

These three rotation gates help the qubit rotate to any desired state from a given input state. 
As a result, starting from state $\ket{0}$, the state of the $i$th qubit after the first layer can be determined by $\ket{\phi_i^1} = \Xi_{i,1}R_Y(r^t_N w_{i,N}) \cdots R_Y(r^t_1 w_{i,1})\ket{0}$, where $\Xi_{i,1}=R_Z(\chi^t_{i,1})R_Y(\rho^t_{i,1})R_Z(\vartheta^t_{i,1})R_Y(r^t_1 w^t_{i,1})\allowbreak R_X(\tilde{v}_t^2) $.
Our multi-layer VQC ansatz is motivated by a data reuploading~\cite{data_reuploading}, which possesses the universal approximation property encouraging that any arbitrary denoising function can be asymptotically approximated. 

Finally, we measure each qubit by a Hermitian observable of the Pauli-Z operator: 
\begin{align}
    P_Z = 
    \begin{bmatrix}
    1 & 0\\ 
    0 & -1 
    \end{bmatrix}
    .
\end{align}
Letting $\ket{\phi_i^L}$ be the state of $i$th qubit after $L$ layers, the expectation value of the observable $P_Z$ is written by:
\begin{align}
    m_i^t = \bra{\phi_i^L}P_Z \ket{\phi_i^L}, 
\end{align}
which is bounded as $-1 \leq m_i^t \leq 1$. 
We adjust the measurement output of VQC to generate the scaling factor as follows:
\begin{align}
    s_{1, i}^t = \frac{1}{2}(m_i^t + 1),
    \label{eq:out}
\end{align}
so that the scaling factor $\frac{s_{1,i}^t}{1+s_{2,i}^t}$ in~(\ref{scaling}) is bounded by $1$ to prevent a possible divergence that might occur during the learning phase of the VQC.
The overall VQC-CS algorithm is summarized in Algorithm~\ref{algo}.

\begin{table}[t]
\begin{algorithm}[H]
\normalsize
\caption{VQC-CS Algorithm}
\label{algo}
\textbf{Input}: Received symbols $\mathbf{y}$ and initial estimate $\hat{\mathbf{x}}^0 = \mathbf{0}$
\textbf{Output}: Non-linear estimate $\hat{\mathbf{x}}^t$ and linear estimate $\mathbf{l}^t$
\begin{enumerate}[\bfseries Step 1:]
\addtolength{\itemindent}{6mm}
    \item Compute the LE estimate $\mathbf{l}^t$ in~\eqref{eq:le}.
    \item Prepare VQC state by residual error in~\eqref{recovery_ee_embed}.
    \item Embed the LE estimate to VQC in~\eqref{r_embedding}.
    \item Measure VQC qubits to obtain scaling factors in \eqref{eq:out}.
    \item Denoise the LE estimate in~\eqref{scaling}.
    \item Repeat Steps 1 through 5 for the preset number of iterations.
\end{enumerate}
\end{algorithm}
\end{table}

\subsection{VQC Training} 

Analogous to deep learning, a cost value of the system is back-propagated to train the VAC. 
Let $T$ denote the total number of compressed sensing iterations. 
We use an exponentially weighted mean-square error loss: 
\begin{align}
    \mathcal{C} = \frac{1}{N} \sum_{t=1}^T \zeta^{T-t}\|
    \hat{\mathbf{x}}^t - \mathbf{x}
    \|^2,
\end{align}
to accumulate the estimation errors of each iteration, where $\zeta$ is a decay factor to adjust the loss contribution of each iteration. 
In consequence, the parameters in the rotation gates are finely tuned by back-propagation during the learning phase. 
Once the rotation angles are determined, the VQC is deployed for testing the performance of compressed sensing in grant-free IoT-device access systems.

\section{Performance Evaluations}

\subsection{System Parameters}

In this section, we present the performance of VQC-CS for joint channel estimation and user identification over a wireless channel with sparse activity in grant-free device access systems. 
We assume Rayleigh fading channel, i.e., channel coefficients $\mathbf{h}$ are samples from $h_i \sim \mathcal{CN}(0, 1/\rho)$, where $\rho$ is a user activity rate.
Without loss of generality, we consider a pilot matrix $\mathbf{A} = \mathbf{\Lambda}\mathbf{V}^\text{H}$, where $\mathbf{\Lambda}$ and $\mathbf{V}$ denote a square diagonal matrix of singular values and corresponding right singular matrix. 
We assume that the summation of singular values is normalized to $N$, i.e.,~$\sum_{i=1}^M\lambda_i=N$, and it holds $\lambda_i/\lambda_{i+1} = \kappa^{1/M}$ for $i \in \{1, 2, \ldots, M-1\}$. 
$\kappa$ indicates the condition number of the pilot matrix and is set to $1$ in the following simulations for simplicity.
The right singular matrix is modeled as $\mathbf{V}^\text{H} = \mathbf{\Pi}\mathbf{F}$, where $\mathbf{\Pi}$ and $\mathbf{F}$ indicate a random permutation matrix and discrete Fourier transform matrix, respectively. 
We consider a user activity rate of $\rho=0.2$, a correlation coefficient of $\gamma=0.6$, and a channel SNR of $30$~dB.

We use pseudo-inverse matrix $\hat{\mathbf{D}}_{\text{PINV}}$ in the LE step for OAMP and VQC-CS algorithms.
The training hyper-parameters of the VQC-CS algorithm are listed in Table~\ref{hyperparameters}.
We use at least $5{,}000$ sample realizations for performance evaluations.

\begin{table}[t]
	\centering
	\caption{Hyper-parameters of training VQC-CS}
	\label{hyperparameters}
		\begin{tabular}{c c}
			\hline 
			Parameters & Values \\
			\hline
			Exponential decay $\zeta$   & $0.85$ \\
			Learning rate & $0.01$ \\
			Number of layers $L$ & $3$ \\
			Optimizer & Root mean square propagation\\
			\hline	
		\end{tabular}
\end{table}

\subsection{Channel State Acquisition}

\begin{figure}[t]
    \centering
    \includegraphics[width=\linewidth]{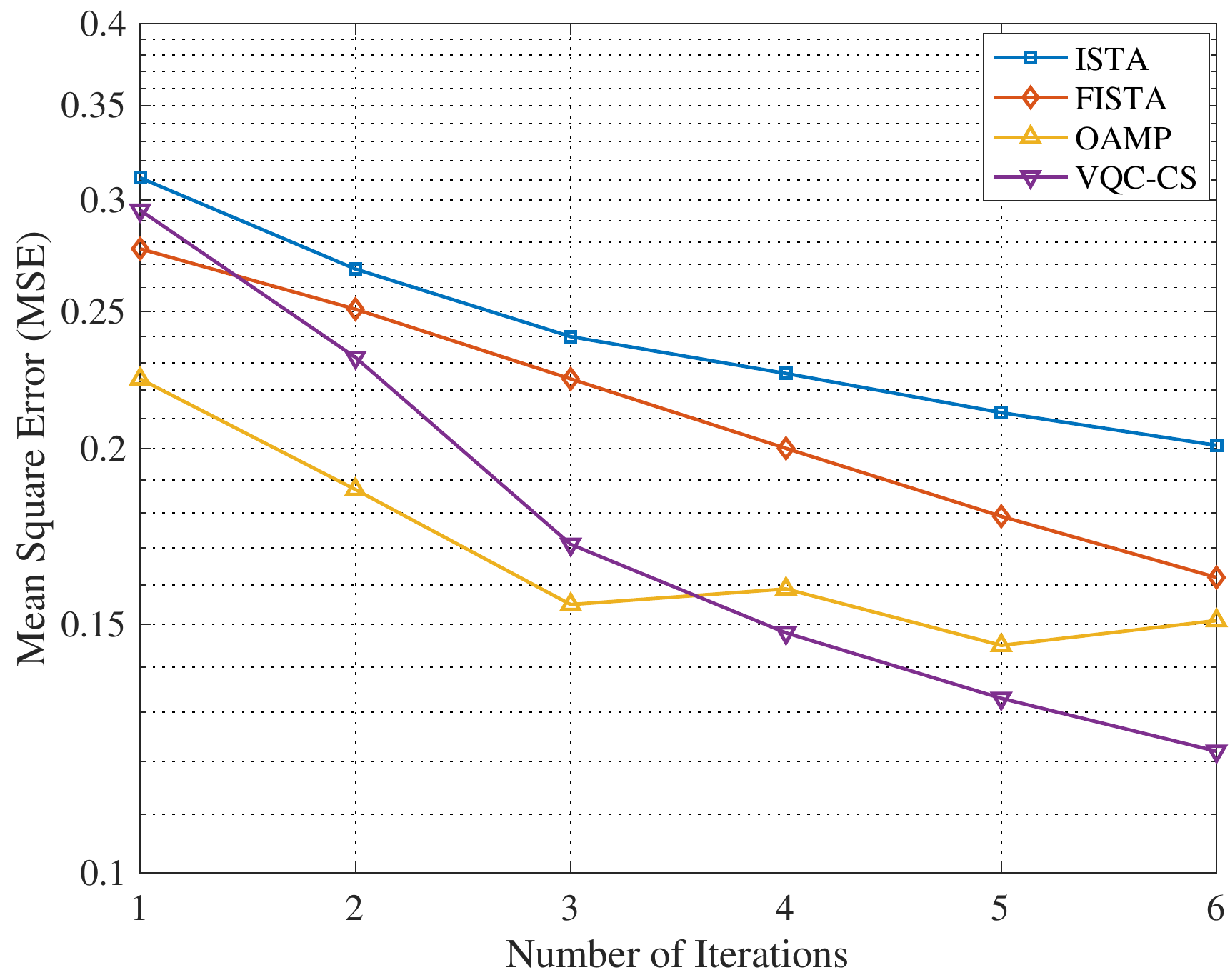}
    \caption{MSE performance across CS iterations for ISTA, FISTA, OAMP and VQC-CS algorithms in 
    systems having $N=10$ devices, $M=7$ received symbols and correlation $\gamma=0.6$.    }
    \label{N_10_M_7_corr06}
\end{figure}

In Fig.~\ref{N_10_M_7_corr06}, we present the MSE performance of CS algorithms for the network system with $N = 10$ devices and $M = 7$ received symbols. 
Note that these algorithms all use an element-wise denoiser function: OAMP employs an MMSE denoiser; and ISTA/FISTA employs a soft-thresholding.
We observe that OAMP outperforms the ISTA and FISTA algorithms as expected. 
After $3$ iterations, VQC-CS shows the best performance in channel estimation among these algorithms.
This may be because the VQC-CS can explore the correlation structure in the user activity.

\begin{figure}[t]
    \centering
    \includegraphics[width=\linewidth]{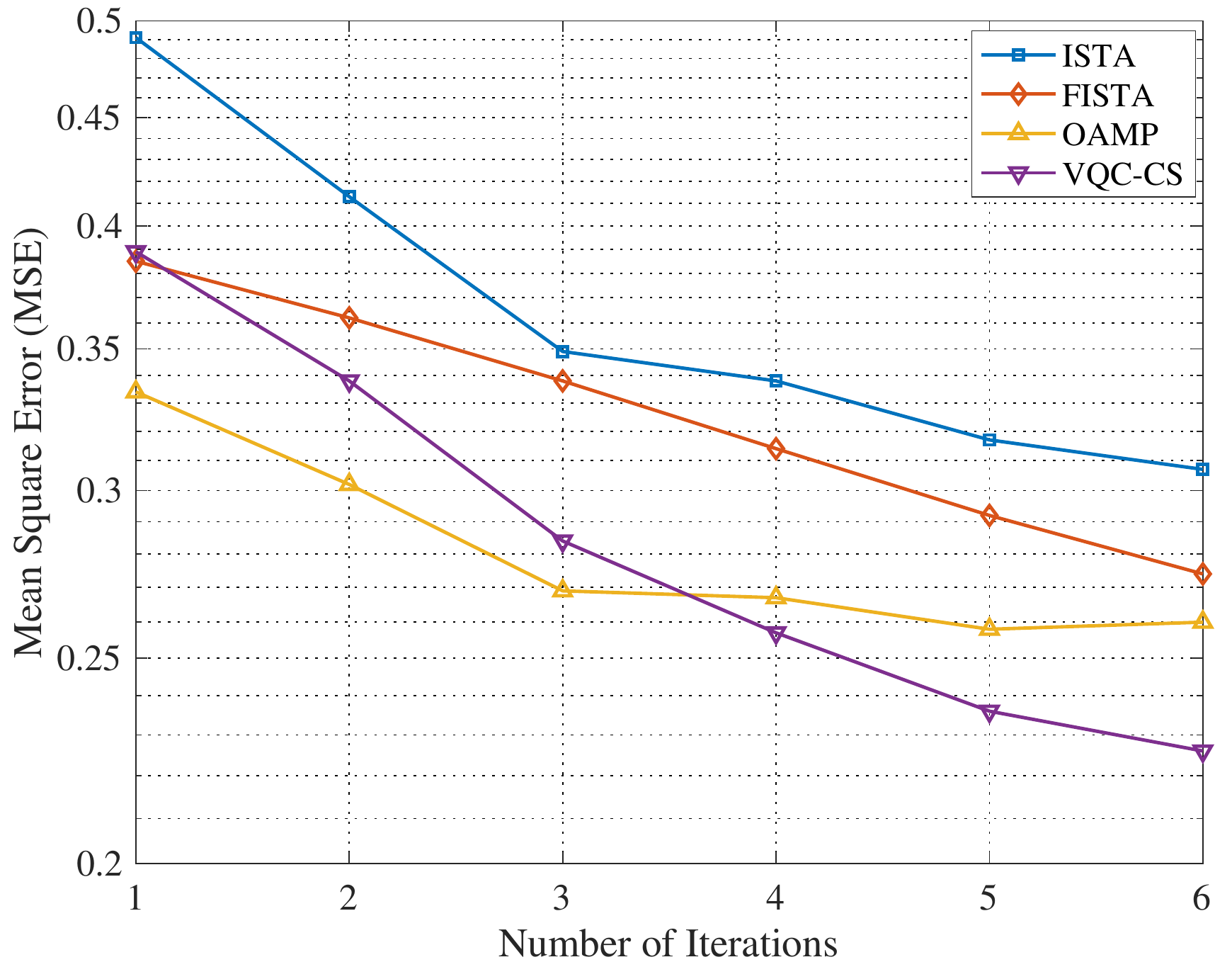}
    \caption{MSE performance across CS iterations for ISTA, FISTA, OAMP and VQC-CS algorithms in systems having $N=10$ devices, $M=6$ received symbols and correlation $\gamma=0.6$. }
    \label{N_10_M_6_DFT_corr_06}
\end{figure}

Fig.~\ref{N_10_M_6_DFT_corr_06} shows the case with fewer received symbols for $N = 10$ and $M = 6$ to evaluate the performance of VQC-CS under a more severe channel condition. 
Similar to the previous case, it is observed that VQC-CS can achieve the best channel estimation performance after $3$ iterations. 
OAMP saturates to a steady-state performance in an early iteration, but the achieved MSE is worse compared to the proposed VQC-CS algorithm.

\subsection{Device Activity Detection}

\begin{figure}[t]
    \centering
    \includegraphics[width=\linewidth]{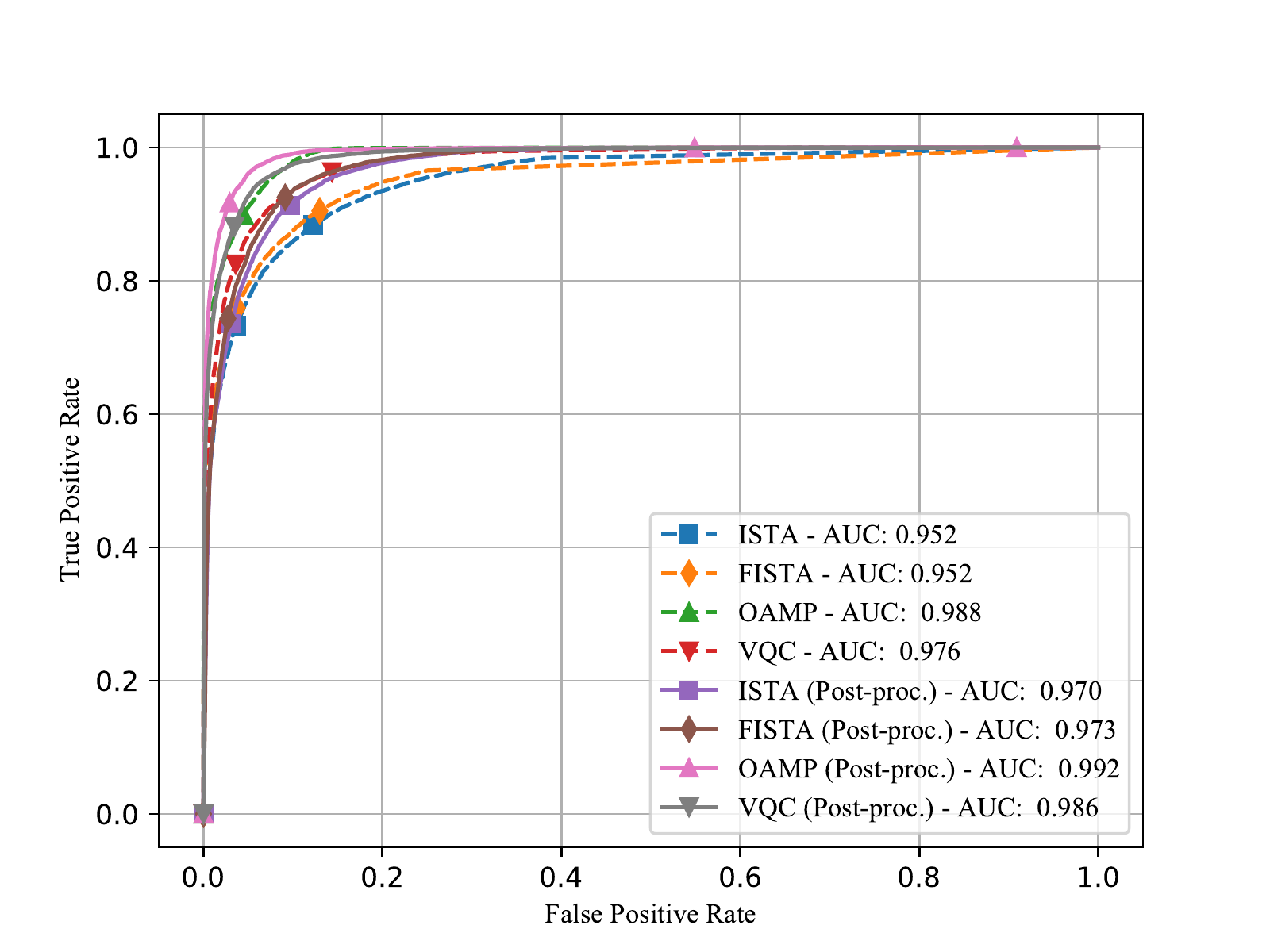}
    \caption{ROC charts for ISTA, FISTA, OAMP and VQC-CS algorithms in 
    systems having $N=10$ devices, $M=6$ received symbols and correlation $\gamma=0.6$.    }
    \label{ROC_graph}
\end{figure}

The CS algorithms can detect the active devices at the same time of channel estimation. 
We evaluate the user identification performance in terms of the area under the curve (AUC) of the receiver operating characteristic (ROC).
The true positive rate and false positive rate are calculated based on thresholding the NLE output, where the device is determined as active if $|\hat{x}^T_i|$ is greater than a threshold.
As an example, \textcolor{black}{we show the AUC of ROC} in Fig.~\ref{ROC_graph} for systems with $N=10$ and $M=6$.
We observe that VQC-CS achieves an AUC-ROC of $0.976$, which is better than the ISTA and FISTA. 

Besides the direct threshold method, we use a post-processing multi-layer perception (two hidden layers with $4N$ and $2N$ neurons respectively) to optimize the binary cross entropy loss for user activity recognition after the CS-based channel estimation. 
The input of the post-processing neural network is $|\hat{x}_i^T|$ for $i \in \{1, 2, \ldots, N\}$.
With the post-processing, the AUC-ROC increased from $0.976$ to $0.986$. 
Since VQC-CS uses MSE as an objective function and the NLE may not follow the orthogonality, the user identification performance of VQC-CS is slightly worse than the OAMP. 
Nevertheless, it was verified that the VQC-CS can outperform ISTA and FISTA both in MSE and AUC scores.

\color{black}
\section{Conclusion}
In this paper, we proposed a new CS algorithm based on VQC, that can be applied to joint channel estimation and user identification in grant-free IoT-device access systems.
The proposed framework is a hybrid classical-quantum computing paradigm, where the NLE step exploits a trainable VQC processor to properly refine the estimate of the LE step as an alternative denoiser. 
We showed that VQC-CS can outperform conventional CS techniques under a challenging system scenario where the device activity is correlated.

This paper is the very initial proof-of-concept study using a variational quantum computing in the area of CS for the future quantum-ready society. 
Note that the quantum processors may not be necessarily better than classical processors in term of prediction accuracy, but  potentially more computationally efficient. 
This new framework has huge potentials beyond just wireless systems.
There remain many fascinating challenges for future work, including rigorous performance verification with real quantum processors and quantum ansatz design for large-scale CS problems.

\bibliographystyle{IEEEtran}
\bibliography{vqc_refs}

\end{document}